\documentstyle[12pt,epsfig]{article}

\newcommand{\be}{\begin{equation}}
\newcommand{\ee}{\end{equation}}
\newcommand{\ba}{\begin{eqnarray}}
\newcommand{\ea}{\end{eqnarray}}

\def\e+e-{$e^+e^-$}               

\def\Nbar{\bar N}               

\def\NF{{\cal N}_{\kern -1.9pt f}}     
\def\NC{{\cal N}_{\kern -1.7pt c}}     

\def\pt{p\kern -.2pt\lower 4pt\hbox{\tiny T}}    

\newcommand{\eref}[1]{(\ref{#1})}      

\def\nostrocostrutto#1\over#2{\mathrel{\mathop{\kern 0pt \rlap
  {\raise.2ex\hbox{$#1$}}}
  \lower.9ex\hbox{\kern-.190em $#2$}}}

\def\p0{P_0(\Delta y)}

\def\Dy{\Delta y}

\def\psigma{{p\kern -.2pt\lower 4pt\hbox{$\scriptstyle\Sigma$}}}
\def\rsigma{{r\kern -.2pt\lower 4pt\hbox{$\scriptstyle\Sigma$}}}

\def\la{\log 2}
\def\yc{y_c}

\def\zmin{{z_{\rm min}}}
\def\zmax{{z_{\rm max}}}

\begin{document}

\title{
\vspace{-3.0cm}
\hspace*{\fill}{\normalsize DFTT 95-32} \\ \vspace{-0.4cm}
\hspace*{\fill}{\normalsize MPI-PhT/95-43} \\ \vspace{-0.4cm}
\hspace*{\fill}{\normalsize LU TP 95-11} \\ \vspace{-0.4cm}
\hspace*{\fill}{\normalsize 20 May 1995} \\*[2.0ex]
{The average number of partons per clan in rapidity intervals in
parton showers}\thanks{Work supported in part by M.U.R.S.T. under Grant 1994}
}

\author{A.\  Giovannini$^1$ \thanks{E-mail: giovannini@to.infn.it}\ , \
 S.\  Lupia$^2$ \thanks{E-mail: lupia@mppmu.mpg.de}\ , \
 R.\ Ugoccioni$^3$ \thanks{E-mail: roberto@thep.lu.se}}

\date{$^1$ {\normalsize\it
Dip. Fisica Teorica and I.N.F.N. -- Sezione di Torino, \\
via Giuria 1, I-10125 Torino, Italy} \\ \vspace{0.2cm}
$^2$  {\normalsize\it
Max-Planck-Institut f\"ur Physik, Werner-Heisenberg-Institut \\
F\"ohringer Ring 6, D-80805 M\"unchen, Germany} \\ \vspace{0.2cm}
$^3$ {\normalsize\it
Dept. of Theoretical Physics, University of Lund \\
S\"olvegatan 14 A, S-223 62 Lund, Sweden} }

\maketitle

\begin{abstract}
The dependence of the average number of partons per clan
on virtuality and rapidity variables is analytically predicted in the
framework of the Generalized Simplified Parton Shower model, based on
the idea that clans are genuine elementary subprocesses.
The obtained results are found to be qualitatively consistent with
experimental trends. This study extends previous results
on the behavior of the average number of clans in virtuality and rapidity
and shows how important physical quantities can be calculated analytically
in a model based on essentials of QCD  allowing
local violations of the energy-momentum conservation law,
still requiring its global validity.
\end{abstract}

\vfill\eject

\section{Introduction}

The first motivation of our search lies
 in the relevance of clan structure analysis both in
theoretical and experimental study of multiparticle production.
It is indeed remarkable that
clan structure analysis of charged particle
Multiplicity Distributions (MD's) in different reactions
and at different c.m. energies,
in full phase space (fps) and in symmetric
rapidity intervals, $\Dy$, revealed an extraordinary series of new
regularities\cite{Schmitz}, whose interpretation is still in part
puzzling.
On the theory side, it is to be pointed out how the idea of the
production of
independent intermediate objects (clans, clusters, strings...)
which then decay following a cascading mechanism, firstly overlooked,
starting from  1986  became slowly
the corner stone of any model of multiparticle production.
The second motivation  lies in
the lack of a sound justification of clan parameters behavior from first
principles within a parton shower model, which is the natural framework of the
present study  of multiparticle production phenomena
in view of the application of
Generalized Local Parton-Hadron Duality (GLPHD)\cite{GLPHD}
as hadronization prescription (see for instance \cite{Ochsvietri}).

Along this line, we introduced  the Generalized Simplified Parton Shower
(GSPS) model,  based on essentials of QCD
and local weakening of conservation laws, and, in this model,
we calculated  analytically\cite{GSPS} the average number of clans produced in
a given rapidity interval $\Dy$
by an ancestor parton of maximum allowed virtuality $W$, $\bar N(\Dy,W)$.
It has been found that $\Nbar(\Dy,W)$ behavior
predicted by the GSPS model is qualitatively consistent
with experimental findings in $e^+e^-$ annihilation
at hadron level. In particular,
$\Nbar(\Dy,W)$ grows in rapidity in a way very close to linear
and then bends toward a constant value as $\Dy$ approaches
full phase space.
$\Nbar(\Dy,W)$ shows also an approximate (5\%) energy independence at
ancestor energies below 100 GeV. This is actually a very slow
decrease which is reminiscent of the results obtained
in Monte Carlo simulations for single gluon jets\cite{single}.
This model shows also the correct growing of
$\Nbar({\mathrm{fps}},W)$ with initial virtuality $W$.
In addition to these results, a new regularity for rescaled quantities
is predicted: the ratio of $\Nbar(\Dy,W)$ to the value
in full phase space $\Nbar({\mathrm{fps}},W)$ is
approximately energy independent
within a much higher degree of approximation than that previously
seen for $\Nbar(\Dy,W)$,
when it is plotted  as a function of rapidity
interval rescaled to full phase space.

These trends of the average number of clans in virtuality and
rapidity variables are really remarkable and
strongly demand to complete the research program
announced at the end of our previous work, i.e.,
to  calculate analytically also
the average number of partons per clan in the same variables,
$\bar n_c(\Dy,W)$, and to check its behavior with experimental data.
In order to do that, first we calculate analytically
the average number of partons in the shower as a
function of rapidity interval $\Dy$ and virtuality $W$;
then, by using previous findings\cite{GSPS}
on the average number of clans, $\Nbar(\Dy,W)$,
we determine the average number of partons per clan, $\bar n_c(\Dy,W)$.

The plan of the paper goes as follows.
In Section~2 clan definition is reexamined in the framework of a two step
process in view of the relevance of multiplicity channels  0 and 1 in our
analysis. In Section~3 a summary of the GSPS model is
presented. In Section~4 the explicit analytical calculation of the average
number of partons in the shower and  per clan in the GSPS model is performed.
Comments on the obtained results are given in Section~5.

\section{Two step processes and clan definition}

As discussed in the Introduction,
the main conclusion of experimental analysis of clan structure parameters
is that the parton production process within
a shower is a two step process: to the initial production of independent
intermediate parton sources (clans), it follows in the second step the
production of partons inside  each clan, whose average number calculation
is the main subject of this paper.

Following
the standard procedure\cite{pzero}, let us assume to produce in the first
step (independently from the second one) $N$ objects ($N = 0,1,\dots$) with
probability $p_N$ and generating function
\be
f(z) = \sum_{N=0}^{\infty} p_N z^N \; .
\ee
Each of the $N$ produced objects gives origin in the second step to
partons according to  the same multiplicity distribution,
$q_{n_i}$, ($n_i = 0,1 \dots$, $\sum_{i= 1}^N n_i = n$), whose
generating function is
\be
g(z) = \sum_{n_i=0}^{\infty} q_{n_i} z^{n_i} \; .
\ee
The two step nature of the process is summarized in the equation:
\be
F(z) \equiv \sum_{n=0}^{\infty} P_n z^n = \sum_{N=0}^{\infty} p_N [g(z)]^N =
f[g(z)] \; ,
\label{ex3}
\ee
where $P_n$ is the probability to produce $n$ partons and $F(z)$ is the
corresponding generating function.
It should be pointed out that Eq.~\eref{ex3} is not limited
to the case in which all produced clans are identical,
but it requires only that all clans can be described by
the same generating function. Suppose in fact that a clan's
MD, $q_{n_i}(\xi)$, depends on a set of parameters, denoted
collectively by $\xi$: if for each clan these parameters
are independent of the values of other clans' parameters and
of the number of clans generated, then it is possible
to define an ``average clan'' whose MD is the MD of a single
clan averaged with the probability distribution function
$\phi(\xi)$ that a clan is produced with parameters $\xi$:
\be
   \tilde q_{n_i} = \int q_{n_i}(\xi) \phi(\xi) d\xi \; .
\label{ex4}
\ee
In this case Eq.~\eref{ex3} would still be valid with the function
$g(z)$ generating now the distribution $\tilde q_{n_i}$.
This is the type of average that will be performed in the
context of the GSPS model in Section 4.1.

Notice that when $f(z)$ is the generating function of a Poissonian
distribution, i.e.:
\be
f(z) = \exp \left[ \bar N(z-1) \right] \; ,
\ee
it follows:
\be
F(z) = \exp \left[ \bar N[g(z)-1] \right] \; .
\ee
These equations fully define the class of Compound Poisson Distributions
(CPD) with all important related properties\cite{voids} and, in
particular, the clan concept as group of particles of common
ancestor\cite{AGLVH:1}.
Notice that each clan contains at least one particle in order to be
uniquely defined.
This fact corresponds to the request that partons'  MD inside
a clan is shifted, i.e.,  $q_0 = 0$.
This remark is of particular interest for the extension of our
discussion on clan parameters from full phase space to rapidity
intervals. In fact, while in full phase space we are sure that each clan
satisfies the condition $q_0 = 0$, in rapidity
intervals one should be careful because if all partons belonging to a clan
fall outside the interval $\Dy$,  then $q_0(\Dy) \not= 0$.
The problem now is how to define clan concept in this case, i.e., how to
{\it redefine}
clan structure parameters with at least one parton inside each
clan in rapidity intervals.
The goal for a CPD is obtained by solving the equation:
\be
\exp \left\{ \bar N({\mathrm{fps}},W) [g_{\Dy}(z) -1] \right\} =
\exp \left\{ \bar N'(\Dy,W) [g'_{\Dy}(z) -1] \right\} \; ,
\label{resc:1}
\ee
where
\be
g_{\Dy}(z=0) \equiv q_0(\Dy) \not=0
\ee
and
\be
g'_{\Dy}(z=0) \equiv q'_0(\Dy) =0 \; .
\ee
Notice that the right-hand side of Eq.~\eref{resc:1} corresponds to the
standard parametrization commonly used in fitting procedures.

{}From Eq.~\eref{resc:1}, one has:
\be
\bar N'(\Dy,W) = \bar N({\mathrm{fps}},W)[ 1 - q_0(\Dy)] = - \log P_0(\Dy) \; ,
\label{nmedioprimo}
\ee
$P_0(\Dy)$ being the rapidity gap probability and
\be
g'_{\Dy}(z) = \frac{g_{\Dy}(z) - q_0(\Dy)}{1 - q_0(\Dy)}
\label{gprimonoanc}
\ee
the rescaled generating function of partons' MD inside a clan.
The average number of partons per clan is given then by:
\be
\bar n'_c(\Dy,W) = \frac{\bar n_c(\Dy,W)}{1 - q_0(\Dy)}
= \frac{\bar N({\mathrm{fps}},W)}{\bar N'(\Dy,W)} \bar n_c(\Dy,W) \; .
\label{ncprimo:noanc}
\ee

It should be noticed that $[1-q_0(\Dy)]$ is the probability that at least
one parton belonging to a clan falls in the rapidity interval $\Dy$, and
$\bar N'(\Dy,W)$ can then be interpreted as the average number of clans in
$\Dy$.
The probability to have $N'$ clans in the
symmetric rapidity interval $\Dy$, $p_{N'}(\Dy,W)$, can be related\cite{GSPS}
 to the corresponding
probability in full phase space, $p_N({\mathrm{fps}},W)$, by means of the
following equation:
\be
p_{N'}(\Dy,W) = \sum_{N=N'}^{\infty} \Pi(N', \Dy|N,{\mathrm{fps}})
p_N({\mathrm{fps}}, W) \; ,
\label{binomial}
\ee
where $\Pi(N', \Dy|N,{\mathrm{fps}})$ is the conditional probability to
have $N'$ clans in $\Dy$ when one has $N$ clans in full phase space.
Since in \cite{GSPS} clans in full phase space are independently produced,
$\Pi(N', \Dy|N,{\mathrm{fps}})$
turns out to be in this case a binomial distribution:
\be
\Pi(N', \Dy|N,{\mathrm{fps}}) = {N \choose N'} \pi(\Dy,W)^{N'}
[1-\pi(\Dy,W)]^{N-N'} \; ,
\ee
where $\pi(\Dy,W)$ is the conditional probability to
have one clan in $\Dy$ when one clan is produced in full phase space.

This result coincides with the previous one
(see Eq.~\eref{nmedioprimo}); one should simply identify
$\pi(\Dy,W)$ with $[1-q_0(\Dy)]$, i.e., with the probability that at least
one parton in a clan falls in $\Dy$.

The discussion so far has dealt with a distribution for clans that allows
events in which no clans are produced, i.e., empty events. While this
may be appropriate at charged hadron level, at parton level it is not
because, in order to have a single jet, we must have at least one
parton, hence at least one clan
(in the following we will refer to this problem as to the
``contribution of the ancestor'' because it is the ancestor parton
which is always present in the jet).
In full phase space, if additional clans are again produced according
to a Poisson distribution, this contribution leads
to the following equation for generating functions:
\be
F_{\mathrm{fps}}(z) = g_{\mathrm{fps}}(z)
\exp \left\{ \lambda({\mathrm{fps}},W) [g_{\mathrm{fps}}(z)-1] \right\} \; ,
\label{poisson:1}
\ee
with
\be
\lambda({\mathrm{fps}},W) = \bar N({\mathrm{fps}},W) -1
\ee
and
\be
g_{\mathrm{fps}}(z=0) \equiv q_0({\mathrm{fps}}) = 0 \; .
\label{poisson:2}
\ee
Accordingly, in rapidity intervals $\Dy$, Eqs.~\eref{poisson:1} and
\eref{poisson:2} become:
\be
F_{\Dy}(z) = g_{\Dy}(z) \exp \left\{ \lambda({\mathrm{fps}},W)
[g_{\Dy}(z)-1] \right\} \; ,
\ee
with
\be
g_{\Dy}(z=0) \equiv q_0(\Dy) \not= 0 \; .
\ee
In order to properly define clans in rapidity intervals $\Dy$,
the calculation goes as in the
previous scheme without the contribution of the ancestor
and the {\it redefinition} procedure gives:
\be
g_{\Dy}(z) \exp \left\{ \lambda({\mathrm{fps}},W)
[g_{\Dy}(z) -1] \right\} =
\exp \left\{ \lambda'(\Dy,W) [g'_{\Dy}(z) -1] \right\}
\ee
with $g'_{\Dy}(0) = 0$.

Notice that the distribution on the r.h.s.\ must now allow for
intervals empty of partons, hence it is not shifted, and also that
the average number of Poisson distributed objects from this
form results to be $\lambda'(\Dy, W)$.
It follows:
\be
\lambda'(\Dy,W) = \lambda({\mathrm{fps}},W) [ 1 - q_0(\Dy)] - \log q_0(\Dy)
= - \log P_0(\Dy)  \; ,
\label{lambdaprimo}
\ee
\be
g'_{\Dy}(z) = 1 + \frac{\lambda({\mathrm{fps}},W)
[g_{\Dy}(z) - 1] + \log g_{\Dy}(z)}{\lambda'(\Dy,W)}
\label{gprimo}
\ee
and
\be
\bar n'_c(\Dy,W) =
\frac{\lambda({\mathrm{fps}},W) + 1}{\lambda'(\Dy,W)} \; \;  \bar n_c(\Dy,W)
\; .
\label{ncprimo:anc}
\ee
Equations~\eref{lambdaprimo}, ~\eref{gprimo} and ~\eref{ncprimo:anc}
should be compared with
Eqs.~\eref{nmedioprimo}, \eref{gprimonoanc} and \eref{ncprimo:noanc},
where the contribution of the
ancestor is not taken into account.

It should be pointed out that the above procedure is correct
 only if the rescaled generating function
of partons' MD inside a clan, $g'_{\Dy}(z)$, is properly defined, i.e., if
\be
g'(z=1) = 1
\ee
and
\be
q'_n(\Dy) \propto \left( \frac{d^ng'_{\Dy}(z)}{dz^n} \right)_{z=0} \ge 0 \; .
\label{ex20}
\ee
Being $g(z=1) = 1$,
thanks to Eq.~\eref{gprimo}, the first condition is trivially fulfilled;
the second condition, Eq.~\eref{ex20}, can be reformulated as follows:
\be
q'_n(\Dy) = \frac{\lambda({\mathrm{fps}},W)}{\lambda'(\Dy,W)} q_n(\Dy)  +
\frac{1}{\lambda'(\Dy,W)}
\left( \frac{d^n \log g_{\Dy}(z)}{dz^n} \right)_{z=0} \ge 0
\ee
and it is fulfilled if
\be
\left( \frac{d^n \log g_{\Dy}(z)}{dz^n} \right)_{z=0} \ge 0
\qquad \forall n \; ,
\label{ex23}
\ee
i.e., if $\log g_{\Dy}(z)$ is a
generating function up to a normalization factor.
Being $\log (g_{\Dy}(z)/q_0(\Dy))$ the
generating function of combinants associated in $\Dy$
with  partons' MD inside a clan\cite{voids},
$g_{\Dy}(z)$, the validity of Eq.~\eref{ex23} for all $n$ implies that
combinants are positive definite, i.e., that the corresponding
MD inside a clan is also a CPD~\cite{combinants}.

Notice that now the binomial convolution of Eq.~\eref{binomial}, applied
in \cite{GSPS}, gives a result different from Eq.~\eref{lambdaprimo}:
\be
 \bar N''(\Dy,W) = [ 1 - q_0(\Dy)] \bar N({\mathrm{fps}},W)
\not= \lambda'(\Dy,W) \; ;
\label{nmediosecondo}
\ee
it is $\bar N''(\Dy,W)$ which should properly be identified as
the average number of clans that produce at least one parton in
the interval $\Dy$.
Thus in the present case in which at least
one clan is always produced one cannot simply identify the
parameter which appears in a CPD description of the MD in a rapidity
interval with the average number of clans actually contributing
to that interval: the former, given in Eq.~\eref{lambdaprimo} diverges when
the interval $\Dy$ goes to fps; the latter, given in Eq.~\eref{nmediosecondo},
 tends to $\bar N({\mathrm{fps}},W)$ in the same limit.
 This point does not arise in the experimental measure
of charged hadron MD for the simple reason that events with no charged
particles in full phase space are allowed (although they are not
measured).  The detailed analysis we are carrying out in this
paper has brought this problem to the surface: earlier work on
the MD at parton level\cite{SPS} has by-passed this difficulty
by neglecting multiplicities 0 and 1 in full phase space: this
was indeed justified as the interest was mainly in small rapidity
intervals and, more important, the concept of clans was still a
statistical one. But in making the step and treating clans as
genuine subprocesses, this resolution is no longer appropriate;
the consequences of this fact
will be discussed in Section 5 together with our results.

Having clarified this important point on clan
definition in a two step process, we can now
proceed to calculate in the framework of the GSPS model
the average number of particles inside a clan as a function
of rapidity and virtuality variables, i.e.:
\be
\bar n_c(\Dy,W) = \frac{dg_{\Dy}(z)}{dz} \biggl|_{z=1} \; .
\label{ex26}
\ee
It should be clear that clan parameters used in experimental analysis
differ from those calculated in the GSPS model via Eq.~\eref{ex26}. In fact,
in experimental analysis, one determines
$g'_{\Dy}(z)$ (Eq.~\eref{gprimo}) and, accordingly,
\be
\bar n'_c(\Dy,W)  = \frac{dg'_{\Dy}(z)}{dz} \biggl|_{z=1} \; .
\label{ex27}
\ee
The primed quantities of Eq.~\eref{ex27} are of course linked to
the unprimed ones of Eq.~\eref{ex26}
by Eqs.~\eref{lambdaprimo}--~\eref{ncprimo:anc}.

It is to be pointed out that the average number of particles in a
given rapidity interval $\Dy$, $\bar n(\Dy,W)$, can be determined
both before and after the redefinition process. One has:
\be
\bar n(\Dy,W)  =
\bar N({\mathrm{fps}},W) \bar n_c(\Dy,W)
\label{ex31}
\ee
and
\be
\bar n(\Dy,W)  = \lambda'(\Dy,W) \bar n'_c(\Dy,W)
\ee
respectively; notice that, as expected, $\bar n(\Dy,W)$
is independent of the redefinition procedure.

\section{The GSPS model: a summary}

The description of a parton shower initiated by an ancestor parton
of virtuality $Q$ splitting into two
partons of virtualities $Q_0$ and $Q_1$ in terms of virtuality and
rapidity variables is usually given within LLA QCD
 evolution equations\cite{GLAP} by:
\be
  {dP_{Q\to Q_0Q_1} \over dt} = {\alpha_s(t) \over 2\pi}
  \int\limits_{\zmin}^{\zmax} P_{Q\to Q_0Q_1}(z_0) dz_0 \; ,
\label{AP}
\ee
where
\be
  \alpha_s(t) = {12\pi \over (11 N_c - 2 N_f) t} \ \ ,
  \qquad \qquad t = \log {Q^2 \over \Lambda^2}\ \ .
\ee
$N_c$ and $N_f$ are the number of colors and flavors respectively and
$\Lambda$ is the QCD scale.

$P_{Q\to Q_0Q_1} (z)$ is the splitting kernel
and $z_0$ is the energy fraction carried away by the parton of virtuality
$Q_0$.  The virtuality at which
the parent parton splits as well as
the energy fraction carried away by the daughter partons
is determined at each splitting.
The main problem in the standard approach concerns how to take care of
energy-momentum conservation law ($z_0 + z_1 = 1$).
In the GSPS model, a different approach is followed.
The branching process is described here in fact in terms of
finite splitting functions\cite{kajantie} and this procedure allows to
fix kinematical limits on the energy fraction $z_0$ by
two-body kinematics\cite{SPS}.

For the virtuality dependence of the splitting function, a
factorized form inspired to QCD
and suitably normalized by a Sudakov form factor term has been chosen:
\be
p_A(Q_0|Q) dQ_0 = p_0^A(Q) C_A(Q) dQ_0 = C_A(Q) {d \over dQ_0}
\left( {1 \over C_A(Q_0)} \right) dQ_0 =
d\left({\log Q_0\over \log Q}\right)^A \; ,
\label{psplit}
\ee
where $A$ at this level of investigation
is the only free parameter of the model and controls the
length of the cascade.

We include next clans as  independently
produced intermediate parton sources  and modify accordingly the
kinematics of the process by allowing  local
violations of the energy-momentum conservation law, still requiring
its global validity, i.e.,
offspring partons of virtualities $Q_i$ can fluctuate according to:
\be
Q_0 + Q_1 \not\le Q \ \ , \quad 1 \ \hbox{\rm GeV}\ \le
Q_i \le Q \quad \hbox{\rm [i=0,1]} \; .
\label{weaken}
\ee
Under  this assumption, the joint probability density
${\cal P}(Q_0 Q_1|Q)$ becomes factorized:
\be
{\cal P}(Q_0, Q_1|Q) dQ_0 dQ_1  = p_A(Q_0|Q) dQ_0 p_A(Q_1|Q) dQ_1 \; .
\label{41}
\ee
Each parton emission is therefore independent in virtuality.
Of course, constraints on the energy fraction carried away by daughter partons
 are also no longer valid, i.e., $z_0+z_1 \not=1$, and
kinematical bounds in rapidity are loosened:
\be
|y_i - y| \le \log {Q \over Q_i} \qquad \hbox{[i=0,1]} \; .
\label{42}
\ee
This new condition modifies the splitting kernel in $z$ which now is
decoupled:
\be
  P(z_0,z_1) dz_0 dz_1 \propto {dz_0 \over z_0}{dz_1 \over z_1} \; .
\label{43}
\ee
By combining Eqs.~\eref{41} and \eref{43}, one sees that the
total bi--dimensional splitting function  is also
decoupled and corresponds to  two independent
emissions of a single parton, i.e.,  in order to describe the
production process, it is enough to follow just one
branch of the shower evolution.
Eq.~\eref{43} can be rewritten in terms of rapidity variables for
each branch:
 \be
Y(|y_i - y_{i+1}|, Q_i, Q_{i+1}) dy_i \propto {dz_i \over z_i}
= dy_i \; ,
\label{y_0}
\ee
where $Y(|y_i - y_{i+1}|,Q_i, Q_{i+1})$ is the probability that a parent
parton with rapidity $y_{i+1}$ generates in a single
step a parton of rapidity $y_i$, being
their virtualities $Q_{i+1}$ and $Q_i$ respectively.
When properly normalized, Eq.~\eref{y_0} turns out to be:
\be
Y(|y_i - y_{i+1}|, Q_i, Q_{i+1})  dy_i = {dy_i \over 2 \log
\left( {Q_{i+1} \over Q_i} \right) }
\theta(\log (Q_{i+1}/Q_i) - |y_i - y_{i+1}|) \; .
\label{gradino}
\ee
By identifying now in each splitting one of the produced partons as the
ancestor parton of a clan, one obtains that clans' production process is
Markoffian and therefore clans are independently produced; notice that
this a
property of the model and not an assumption, as discussed in \cite{GSPS}.
The explicit calculation of
clans' MD in a parton shower originated by an ancestor parton
of virtuality $W$ gives in fact
a shifted poissonian distribution with average number
\be
\bar N({\mathrm{fps}},W) =
1 + A \log \left( {\log W \over \log 2} \right) \; .
\ee
Results of this scheme of parton shower evolution have been
summarized in the Introduction.

We concluded that
the observed qualitative agreement of the predictions of the model with
experimental behavior for the average number of clans in rapidity
intervals supports our idea that clans can be considered
independent intermediate parton sources, i.e., that local
fluctuations in the virtualities of produced partons can occur
along the shower.
This fact is quite acceptable since energy-momentum conservation laws
are expected to weaken in limited rapidity intervals whereas their
r\^ole is surely fundamental in full phase space.

Finally notice also
that the approximation used in the GSPS model closely resembles that
used in the dipole model\cite{lund},
where two subsequent dipole emissions are
needed in order to describe a full branching. Furthermore, the
kinematical structure of the model is similar. The main difference among
the two models lies in the description of the shower evolution: in fact,
it is based on  finite splitting functions in the GSPS model,
whereas it follows the standard approach with
elementary splitting kernels in the dipole model.

\section{The average number of partons per clan in the GSPS model}

\subsection{The structure of the calculation}

In order to study the average number of clans in the rapidity interval
$\Dy$, $\bar N(\Dy,W)$, in \cite{GSPS}
we limited our discussion to the first step of parton shower
evolution in the GSPS model. It is clear that if one wants to
calculate the average number of partons per clan in the same interval,
$\bar n_c(\Dy,W)$, one has to analyze the second step of parton shower
evolution, i.e., to study the production of partons
inside clans.
In order to do that, inspired by the criterium of
simplicity and previous findings\cite{SPS,AGLVH:3}, we decided
to maintain inside a clan
the structure of the model seen in the first step.
The only difference (see Eq.~\eref{psplit}) lies in the
introduction of a new parameter, $a$, controlling the length of the
cascade inside a clan, in the expression of the probability that a parton
of virtuality $Q$ emits a daughter parton in the virtuality range
[$Q_0, Q_0+dQ_0$], i.e.,
\be
p_a(Q_0|Q) dQ_0 = d\left({\log Q_0\over \log Q}\right)^a \; .
\label{psplit2}
\ee
The model becomes of course a two-parameter model ($A$ and $a$
controlling the length of the cascade in step 1 and 2 respectively).

Let us now consider a clan splitting at  virtuality $Q$ and study the
evolution equation for the generating function of partons' MD in full
phase space inside a clan, $g_{\mathrm{fps}}(z,Q)$, which,
according to the pure-birth structure of the model, turns out to be:
\be
\frac{dg_{\mathrm{fps}}(z,Q)}{dQ} = p_a(Q|Q)
\left[ g_{\mathrm{fps}}^2(z,Q ) - g_{\mathrm{fps}}(z,Q ) \right] \; ,
\label{evol}
\ee
where $p_a(Q|Q)$, from Eq.~\eref{psplit2}, is given by
\be
p_a(Q|Q) = \frac{a}{Q \log Q} \; .
\ee
Notice that, as expected, Eq.~\eref{evol} corresponds to the
evolution equation
for the generating function of gluons' MD in a gluon jet in Leading Log
Approximation of QCD with fixed cutoff infrared regularization\cite{QCDAG},
as well as to the limit of the Simplified Parton Shower
model when one weakens conservation
laws according to Eq.~\eref{weaken}\cite{SPS}.
Of course Eq.~\eref{evol} concerns now partons' MD inside a single clan.

The solution of Eq.~\eref{evol} is
\ba
g_{\mathrm{fps}} (z,Q ) &=&
\frac{z}{1 +(\bar n(Q )-1)(1-z)}  \hskip1.0truecm
Q  \ge 2\ {\mathrm{GeV}} \; ; \nonumber \\
                        &=& \hskip2.0truecm  z  \hskip2.7truecm Q
< 2 \ {\mathrm{GeV}} \; ,
\label{furry}
\ea
where
\be
\bar n(Q) = e^{\lambda_a(Q)}\; ,
\qquad \lambda_a(Q) = \int_2^Q p_a(Q'|Q')dQ' \; .
\label{lambda}
\ee
The solutions correspond to a shifted
geometric distribution (Q $\ge $ 2 GeV)
and to a clan with only one
parton (Q $<$ 2 GeV). The bound 2 GeV is a consequence of the fact
that in the GSPS model the virtuality cutoff is fixed at 1 GeV
(a parton with virtuality $Q  < 2 $ GeV cannot split any further
by assumption).
Notice that this finding agrees with the clan model discussed in
\cite{AGLVH:3}, where logarithmic MD for partons inside average clans is
interpreted as the result of an
average over geometrically distributed single clans of different
multiplicity.

Equation~\eref{furry} answers our first question and gives the
generating function of partons' MD inside a single clan in full phase space.
In order to match our theoretical calculations with experiments, one
should introduce the rapidity
dependence in the model and average single clan properties over clans of
different rapidity and virtuality.
According to our idea that single clans are independent
intermediate parton (gluon)
sources acting both in virtuality and rapidity
spaces, and not only statistical objects as in the original clan model, we
decided first to extend the model defined in full phase space to rapidity
intervals and then to average these results on single clans' MD's.

Following the discussion contained in Section~2, we calculate the
generating function of partons' MD inside a clan in rapidity intervals $\Dy$
through a binomial convolution on the corresponding generating function
in full phase space:
\ba
g_{\Dy}(z,Q,y) &=& \frac{1 + (z-1) \pi_a(\Dy,Q,y)}
{1 +\pi_a(\Dy,Q,y)(\bar n(Q)-1)(1-z)}  \hskip0.8truecm
Q  \ge 2 \ {\mathrm{GeV}} \; ; \nonumber \\
               &=& 1 + (z-1) \pi_a(\Dy,Q,y)
 \hskip1.truecm Q  < 2 \ {\mathrm{GeV}} \; ,
\label{furry:Dy}
\ea
where $\pi_a(\Dy,Q,y)$ is the probability that a clan of initial
virtuality $Q$ and rapidity $y$ produces a daughter parton inside the
interval $\Dy$. Notice that
\be
\pi_a(\Dy,Q,y) = \int_{\Dy} \rho(y_f|Qy) dy_f  \; .
\label{pigreca}
\ee
$\rho(y_f|Qy)$ is the conditional probability that a final parton
produced by a clan of given $Q$ and $y$ has rapidity $y_f$.

We have now extended formally the generating function of partons' MD inside
a clan from full phase space to
rapidity intervals for a single clan of known initial virtuality $Q$
and rapidity $y$. It is clear that in this model a single event contains
many clans of different initial parameters, while at experimental level one
measures properties of average clans, in the sense indicated in
Eq.~\eref{ex4}. Accordingly, we proceed now to average
 Eq.~\eref{furry:Dy} over $Q$ and  $y$. It follows that:
\be
\bar g_{\Dy}(z,W) = \int_1^W dQ \int_{- \infty}^{\infty} dy
g_{\Dy}(z,Q,y) \sigma(Qy|W) \; ,
\label{genfun}
\ee
where $\bar g_{\Dy}(z,W)$ is the generating function of partons' MD in
the interval $\Dy$ for an average clan generated in a shower of
virtuality $W$; $\sigma(Q y|W)$ is the probability that a parton of
maximum allowed virtuality $W$ produces a clan of virtuality $Q$ and
rapidity $y$ and $g_{\Dy}(z,Qy)$ is given by Eq.~\eref{furry:Dy}.

By deriving then Eq.~\eref{genfun} with respect to $z$
and choosing $z$=1, one gets
the average number of partons in an
average clan generated in a shower of virtuality $W$ in the rapidity
interval $\Dy$, i.e.:
\be
\bar n_c(\Dy,W) = \int_1^W dQ \int_{- \infty}^{\infty} dy \;
\bar n_c(\Dy,Q,y) \sigma(Q,y|W)  \; ,
\label{nmedio:master}
\ee
where $\bar n_c(\Dy,Q,y)$ is the parton average multiplicity in a
rapidity interval $\Dy$ for a single clan of given $Q$ and $y$,
and is related to $\bar n_c({\mathrm{fps}},Q,y)$ as follows:
\be
\bar n_c(\Dy,Q,y) = \pi_a(\Dy,Q,y)  \bar n_c({\mathrm{fps}},Q,y) =
\pi_a(\Dy,Q,y) e^{\lambda(Q)} \; .
\label{nmedio}
\ee
Equation~\eref{nmedio:master} formally solves our main problem;
the final analytical expression for $\bar n_c(\Dy,W)$ results after
having determined explicitly the two factors $\bar n_c(\Dy,Q,y)$
and $\sigma(Q,y|W)$. Notice that the integration domain in $y$ is
determined by kinematical constraints in $\sigma(Q,y|W)$, which
are explicitly discussed in the following Subsection.

In Subsections~4.2 and 4.3, we determine
$\sigma(Q,y|W)$ and $\bar n_c(\Dy,Q,y)$  respectively, whereas
in Subsection~4.4 we perform the final integration over $Q$ and $y$
according to Eq.~\eref{nmedio:master}.

\subsection{The probability to produce a clan of given virtuality
$Q$ and rapidity $y$, $\sigma(Q,y|W)$ }

The weight function $\sigma(Q,y|W)$ we are looking for coincides with
the bi--dimensional clan density in virtuality $Q$
and rapidity $y$ for a shower originated by a parton of maximum allowed
virtuality $W$, $p_{\Sigma}(Q,y|W)$,  up to the normalization factor
$\bar N({\mathrm{fps}},W)$:
\be
\sigma(Q,y|W) =  \frac{p_{\Sigma}(Q,y|W)}{ \bar N({\mathrm{fps}},W)} \; .
\label{sigma}
\ee
Notice that $p_{\Sigma}(Q,y|W)$ by definition depends on the first step
of the production process and therefore the expression of $\sigma(Q,y|W)$
is expected to contain the parameter $A$ only.

Following the procedure used in \cite{GSPS}, where we calculated the
clan density in rapidity,
we define here $p_{\Sigma}(Q,y|W)$ as:
\be
p_{\Sigma}(Q,y|W) = \sum_{N=1}^{\infty} p_N(Q,y|W) \; ,
\label{psigma}
\ee
where $p_N(Q,y|W)$ is the probability that a parton of maximum allowed
virtuality $W$ generates  a clan of virtuality $Q$ and rapidity $y$
after $N$ steps.

The explicit formula for $p_N(Q,y|W)$ in the GSPS model is given by:
\ba
&p_N(Q y|W)& = \nonumber \\
&\int_{{\mathrm{max}}\{2,Q\}}^W&
dQ_{N-1} p_A(Q_{N-1}|W) \int_{-\infty}^{\infty} dy_{N-1}
\delta(y_{N-1} - {\mathrm{tanh}}^{-1}
\sqrt{1 - (Q_{N-1}/W)^2} ) \nonumber \\
&\int_{{\mathrm{max}}\{2,Q\}}^{Q_{N-1}}& dQ_{N-2} p_A(Q_{N-1}|Q_{N-2})
\int_{-\infty}^{\infty} dy_{N-2} Y(|y_{N-2}-y_{N-1}|,Q_{N-2},Q_{N-1})
 \nonumber \\
&\dots& \\
&\int_{{\mathrm{max}}\{2,Q\}}^{Q_2}& dQ_1 p_A(Q_1|Q_2)
\int_{-\infty}^{\infty} dy_1 Y(|y_1-y_2|,Q_1,Q_2) \nonumber \\
&p_A(Q|Q_1)&  Y(|y-y_1|,Q,Q_1) \; . \nonumber
\ea
Following the GSPS model prescriptions, $Y(|y_i - y_{i+1}|, Q_i, Q_{i+1})
$ can be approximated by a gaussian function, i.e.:
\be
Y(|y_i - y_{i+1}|, Q_i, Q_{i+1})  = { 1 \over \sqrt{2 \pi
\log \left( {Q_{i+1} \over Q_i} \right) }}
\exp \left( - {|y_i - y_{i+1}|^2 \over 2
\log \left( {Q_{i+1} \over Q_i} \right) } \right) \; .
\ee
This approximation allows to perform the integration over all
rapidity variables with the exception of $y_{N-1}$.
The integrations on the corresponding  virtuality variables can be done
by using the symmetry properties of the integrand.
One gets:
\ba
& &p_N(Q y|W) = \nonumber \\
& &\int_Q^W dQ_{N-1} p_A(Q_{N-1}|W)
\int_{-\infty}^{\infty} dy_{N-1}
\delta(y_{N-1} - {\mathrm{tanh}}^{-1}
\sqrt{1 - (Q_{N-1}/W)^2} ) \nonumber \\
& &p_A(Q|Q_{N-1}) Y(|y-y_{N-1}|,Q,Q_{N-1})
\frac{[\lambda_A(Q_{N-1}) - \lambda_A(Q) ]^{N-2}}{(N-2)!}
\nonumber \\
& &{\mathrm{for}} \  \ \  Q \ge 2 \ {\mathrm{GeV}}
\ea
and
\ba
& &p_N(Q y|W) = \nonumber \\
& &\int_2^W  dQ_{N-1} p_A(Q_{N-1}|W)
\int_{-\infty}^{\infty} dy_{N-1}
\delta(y_{N-1} - {\mathrm{tanh}}^{-1}
\sqrt{1 - (Q_{N-1}/W)^2} )\nonumber \\
& &p_A(Q|Q_{N-1}) Y(|y-y_{N-1}|,Q,Q_{N-1})
\frac{[\lambda_A(Q_{N-1})]^{N-2}}{(N-2)!}
\nonumber \\
& & {\mathrm{for}} \  \ \  Q < 2\ {\mathrm{GeV}} \; .
\ea
Notice that the last integration over $y_{N-1}$ needs not to be
performed by using the gaussian approximation; in fact, the integration
can be done  with
$Y(|y_i - y_{i+1}|, Q_i, Q_{i+1})$ given in its original form
(Eq.~\eref{gradino}).
This fact restores global conservation laws in rapidity space.

By using Eq.~\eref{psplit} for $p_A(Q_0|Q)$ and Eq.~\eref{lambda}
for $\lambda_A(Q)$, Eq.~\eref{psigma} leads to the following expression
for $p_{\Sigma}(Q,y|W)$:
\ba
p_{\Sigma}(Q y|W) &=&  p_0^A(Q) C_A(W) \int_Q^W dQ_{N-1} p_0^A(Q_{N-1}) C_A(Q)
\nonumber \\
& &\int_{-\infty}^{\infty} dy_{N-1}
\delta(y_{N-1} - {\mathrm{tanh}}^{-1}
\sqrt{1 - (Q_{N-1}/W)^2} ) Y(|y-y_{N-1}|,Q,Q_{N-1}) \nonumber \\
&+& p_A(Q|W)
\delta(y - {\mathrm{tanh}}^{-1} \sqrt{1 - (Q/W)^2} ) \nonumber \\
& & {\mathrm{for}} \  \ \  Q \ge 2\ {\mathrm{GeV}}
\ea
and
\ba
p_{\Sigma}(Q y|W) &=& p_0^A(Q) C_A(W) \int_2^W
dQ_{N-1} p_0^A(Q_{N-1}) C_A(2)
\nonumber \\
& &\int_{-\infty}^{\infty} dy_{N-1}
\delta(y_{N-1} - {\mathrm{tanh}}^{-1}
\sqrt{1 - (Q_{N-1}/W)^2} ) Y(|y-y_{N-1}|,Q,Q_{N-1}) \nonumber \\
&+& p_A(Q|W)
\delta(y - {\mathrm{tanh}}^{-1} \sqrt{1 - (Q/W)^2}) \nonumber \\
& & {\mathrm{for}} \  \ \  Q < 2\ {\mathrm{GeV}} \; .
\ea
By integrating over $y_{N-1}$, one has:
\ba
p_{\Sigma}(Qy|W) &=& p_A(Q|W) C_A({\mathrm{max}}(2,Q))
\int_{{\mathrm{max}}(2,Q)}^W dQ_{N-1} \frac{p_0^A(Q_{N-1})}
{2 \log Q_{N-1}/Q}
\nonumber \\
& &\theta(\log Q_{N-1}/Q - |y - {\mathrm{tanh}}^{-1}
\sqrt{1 - (Q_{N-1}/W)^2}|) \nonumber \\
&+& p_A(Q|W)
\delta(y - {\mathrm{tanh}}^{-1} \sqrt{1 - (Q_{N-1}/W)^2}) \; .
\label{ex61}
\ea
Equation~\eref{ex61} in the approximation
\be
{\mathrm{tanh}}^{-1} \sqrt{1 - (Q_{N-1}/W)^2}
\simeq \log (2W) - \log Q_{N-1} \; ,
\label{tanh}
\ee
gives:
\ba
p_{\Sigma}(Qy|W) &=& p_A(Q|W) C_A(Q)
\int_{1/2(\log 2W - y + \log Q}^{\log W}
d(\log Q') Q'
\frac{p_0^A(Q')}{2 \log Q'/Q} \nonumber \\
&+& p_A(Q|W)
\delta(y - {\mathrm{tanh}}^{-1} \sqrt{1 - (Q'/W)^2}) \nonumber \\
&{\mathrm{for}}& \  \ \  Q \ge 2\ {\mathrm{GeV}}
\label{psigma:more2}
\ea
and
\ba
p_{\Sigma}(Q y|W) &=& p_A(Q|W) C_A(2)
\nonumber \\
& &\int_{{\mathrm{max}}(2,Q)}^W dQ' \frac{p_0^A(Q')}
{2 \log Q'/Q} \theta(\log Q'/Q - |y - {\mathrm{tanh}}^{-1}
\sqrt{1 - (Q'/W)^2}|) \nonumber \\
&+& p_A(Q|W)
\delta(y - {\mathrm{tanh}}^{-1} \sqrt{1 - (Q'/W)^2} )\nonumber \\
& &{\mathrm{for}} \  \ \  Q < 2\ {\mathrm{GeV}} \; .
\label{psigma:less2}
\ea
The region where $p_{\Sigma}(Q y|W)$ is different from zero, according
to the just mentioned approximation, is given by
\be
- \log \frac{W}{2Q} \le y \le  \log \frac{2W}{Q} \; .
\label{kin:triangle}
\ee
These results allow to calculate $\sigma(Q,y|W)$ from
Eq.~\eref{sigma}.

\subsection{The average
number of partons in $\Dy$ inside a single clan, $\bar n_c(\Dy,Q,y)$}

In view of Eqs.~\eref{pigreca} and \eref{nmedio}, one has now to calculate
the conditional probability that a final parton produced
by a clan of given virtuality $Q$ and rapidity $y$ has rapidity $y_f$,
$\rho(y_f|Qy)$.
This calculation can be performed in terms of the probability that a clan of
given virtuality $Q$ and rapidity $y$ produces a final parton of
rapidity $y_f$ after $N$ steps, $r_N(y_f|Qy)$. It follows:
\be
\rho(y_f|Qy) = \sum_{N=1}^{\infty} r_N(y_f|Qy) \; .
\ee
Notice that $r_N(y_f|Qy)$ is obtained by integrating $r_N(Q_fy_f|Qy)$
over the virtuality of the final parton, $Q_f$ in the interval 1 GeV
$\div$2 GeV:
\be
r_N(y_f|Qy)  \equiv \int_1^2 dQ_f r_N(Q_fy_f|Qy) \; .
\ee
$r_N(Q_fy_f|Qy)$ is in fact the probability that a final parton of virtuality
$Q_f$ and rapidity $y_f$ is generated by a clan of virtuality $Q$ and rapidity
$y$ after $N$ steps.
The problem is now how to calculate $r_N(Q_f y_f|Qy)$.
This calculation can be performed in the  GSPS model; in fact, being
\ba
r_N(Q_fy_f|Q y) &=& \int_2^Q dQ_{N-1} p_a(Q_{N-1}|Q)
\int_{-\infty}^{\infty} dy_{N-1} Y(|y_{N-1}-y|,Q_{N-1},Q)
\nonumber \\
& & \dots \\
& & \int_2^{Q_2} dQ_1 p_a(Q_1|Q_2)
\int_{-\infty}^{\infty} dy_1 Y(|y_1-y_2|,Q_1,Q_2) \nonumber \\
& & p_a(Q_f|Q_1)  Y(|y_f-y_1|,Q_f,Q_1) \; , \nonumber
\ea
by using the same technique discussed in the previous Subsection (Gaussian
approximation of $Y(|y_i- y_{i+1}|,Q_i,Q_{i+1})$), one finds:
\be
r_N(Q_fy_f|Q y) = p_a(Q_f|Q) Y(|y_f-y|,Q_f,Q)
\frac{\lambda(Q)^{N-1}}{(N-1)!} \; .
\ee
The structure of this calculation is very similar to that performed in
\cite{GSPS} (Eq.~(48)). The main difference lies in the fact that here we
are studying the production mechanism for partons inside a clan and not for
clans themselves, i.e.,  the result must depend on the
new parameter $a$ only.

Finally, one should notice that in this scheme the ancestor parton after the
splitting continues on its way whereas the daughter parton gives origin to the
clan. Therefore partons' MD inside the clan is shifted by one unit.

Our calculation can be simplified by assuming $Q_f = 1$ GeV in
 $Y(|y_f-y|,Q_f,Q)$, an assumption which is verified for clans
of high virtuality and is quite acceptable in general.
Accordingly, one has:
\be
\rho(y_f|Qy) = e^{\lambda_a(Q)} \int_1^2 dQ_f p_a(Q_f|Q) Y(|y_f-y|,Q_f,Q)
\ee
which, by using Eqs.~\eref{psplit} and \eref{gradino}, becomes
\be
\rho(y_f|Qy) = \frac{1}{2 \log Q} \theta(\log Q - |y_f - y|) \; .
\label{ex69}
\ee
Equation~\eref{ex69} is valid for $Q \ge 2$ GeV.
For $Q <2$ GeV, one produces  clans with only one parton, i.e.:
\be
\rho_{Q<2}(y_f|Qy) = \delta(y-y_f) \; .
\ee
Having found $\rho_{Q<2}(y_f|Qy)$, we can now calculate from
Eqs.~\eref{nmedio}  and~\eref{pigreca} the average number of partons in a
single clan:
\ba
\bar n_c(\Dy,Q,y) &=&  e^{\lambda_a(Q)}
\max \left[ 0 , \frac{\min ({\mathrm{sup}}(\Dy),
y+ \log Q) - \max({\mathrm{inf}}(\Dy),y-\log Q)}{2 \log Q} \right]
\nonumber \\
&{\mathrm{for}}& \  \ \  Q \ge 2\ {\mathrm{GeV}}
\label{nfurry:more2}
\ea
and
\ba
\bar n_c(\Dy,Q,y) &=&  e^{\lambda_a(Q)}   \qquad y \in \Dy \nonumber \\
                  &=&  0  \qquad y \not\in \Dy \nonumber \\
{\mathrm{for}} \  \ \  Q < 2\ {\mathrm{GeV}} \qquad & &
\label{nfurry:less2}
\ea
respectively.

\subsection{The average number of partons in $\Dy$ in an average clan}

The main content of this Section concerns the explicit analytical
expression of $\bar n_c(\Dy,W)$. This result follows by solving the
double integral over $Q$ and $y$ in Eq.~\eref{nmedio:master};
$\bar n_c(\Dy,Q,y)$ is given here by Eqs.~\eref{nfurry:more2} and
\eref{nfurry:less2}, whereas $\sigma(Q,y|W)$ is obtained by inserting
Eqs.~\eref{psigma:more2} and \eref{psigma:less2} into Eq.~\eref{sigma}.

Attention should be paid to the integration domain of $y$ and $Q$, which
is determined by the conditions 1  GeV $\le Q \le W$ and $- \log W/2Q \le y
\le \log 2W/Q$ (see Eq.~\eref{kin:triangle}).
Notice that according to
Eqs.~\eref{nfurry:more2} and~\eref{nfurry:less2}  the integrand has a
different structure in the intervals 1 GeV $\le Q \le 2$ GeV and
2 GeV $\le Q \le W$ GeV. Accordingly, we perform separately the two
integrations:
\be
\bar n_c(\Dy,W) = \bar n_c(Q < 2) + \bar n_c(Q \ge 2) \; .
\ee

In the interval 1 GeV $\le Q \le 2$ GeV, one has:
\ba
\label{lessdue}
\bar n_c(Q<2) &=&
\int_1^2 dQ \int_{- \log W/2Q}^{\log 2W/Q} dy \theta(y_c-|y|)
p_A(Q|W) C_A(2) \\
& & \int_{\max[\log 2,(\log 2W-y+\log Q)/2]}^{\log W}
d\log Q' Q' \frac{p_0^A(Q')}{2 \log Q'/Q} + \nonumber \\ \nonumber
+ \int_1^2 &dQ& \int_{- \log W/2Q}^{\log 2W/Q} dy \theta(y_c-|y|)
p_A(Q|W) \delta(y - {\mathrm{tanh}}^{-1} \sqrt{1-(Q/W)^2} ) \; .
\ea
In the approximation of Eq.~\eref{tanh}, Eq.~\eref{lessdue} can
be rewritten as follows:
\ba
\bar n_c(Q<2) &=& \int_1^2 dQ p_A(Q|W) C_A(2) \Sigma^A(Q,y,W) + \\
              &+& \theta(y_c - \log W) \left[ \left( \frac{\log 2}{\log W}
                  \right)^A - \left( \frac{\log 2W-y_c}{\log W} \right)^2
                  \right] \; , \nonumber
\ea
where
\ba
& &\Sigma^A(Q,y,W)  \equiv \int_{(\log2W-y+\log Q)/2}^{\log W}
d\log Q' Q' \frac{p_0^A(Q')}{2 \log Q'/Q}  = \nonumber \\
& & \frac{A}{2} (\log Q)^{A-1} \left\{ \log x  + \sum_{n=1}^{\infty}
{A-1 \choose n} \frac{x^n}{n} \right\}^{\log W/\log Q -1}_{(\log W - y +
\log Q)/2 \log Q - 1} \; .
\ea
In particular, for $A$=2, one gets:
\ba
\Sigma^2(Q,y,W) &=& \log Q \left[ \log \left( \frac{\log W}{\log Q} - 1
\right) - \log \left( \frac{\log 2W - y + \log Q}{2 \log Q} - 1 \right)
\right]  +  \nonumber \\
&+& \frac{y}{2} + \frac{1}{2} \log W/2Q \; .
\ea

\medskip

Coming to the interval 2 GeV $\le Q \le W$,
one can split the integration in two different terms:
\be
\bar n_c(Q\ge 2)  = \bar n_c(Q\ge 2; \alpha) + \bar n_c(Q\ge 2; \beta) \; ,
\ee
where
\ba
& &\bar n_c(Q\ge 2;\alpha) =
\int_2^W dQ \int_{- \log W/2Q}^{\log 2W/Q} dy
\frac{e^{\lambda_a(Q)}}{\bar N({\mathrm{fps}},W)} \nonumber \\
& & \max \left[0, \frac{\min ({\mathrm{sup}}(\Dy),
y+ \log Q) - \max({\mathrm{inf}}(\Dy),y-\log Q)}{2 \log Q} \right]
\nonumber \\
& & p_A(Q|W) \delta(y - {\mathrm{tanh}}^{-1} \sqrt{1-(Q/W)^2} )
\label{alfa}
\ea
and
\ba
& & \bar n_c(Q\ge 2;\beta) =
\int_2^W dQ \int_{- \log W/2Q}^{\log 2W/Q} dy
\frac{e^{\lambda_a(Q)}}{\bar N({\mathrm{fps}},W)} \nonumber \\
& & \max \left[0, \frac{\min ({\mathrm{sup}}(\Dy),
y+ \log Q) - \max({\mathrm{inf}}(\Dy),y-\log Q)}{2 \log Q} \right]
\nonumber \\
& & p_A(Q|W) C_A(Q) \Sigma^A(Q,y,W) \; .
\ea
So far the rapidity interval has been generic, but in order to continue our
explicit calculations in agreement with standard practice\cite{GSPS}, we
specialize in the following to central symmetric intervals of rapidity
of half-width $y_c$: $\Dy = [- y_c, y_c]$.

By using Eq.~\eref{tanh}
and the properties of the $\delta$-function, one has:
\ba
\label{alfadue}
\bar n_c(Q\ge 2;\alpha) &=&
\int_2^W dQ \frac{1}{2 \log Q} \frac{ C_a(2) p_0^A(Q) C_A(W)}{C_a(Q)
\bar N({\mathrm{fps}},W)}  \\ \nonumber
& & \max \left[0, \min (y_c, \log 2W) -
\max(-y_c,\log 2W - 2 \log Q) \right]   \; .
\ea
Notice that the expression of the maximum in Eq.~\eref{alfadue}
assumes the following different values in different ranges of $\log Q$, i.e.,
$0$ for $\log Q < \frac{1}{2} (\log 2W - y_c)$,
$y_c - \log 2W + 2 \log Q$ for $\frac{1}{2} (\log 2W - y_c) \le
\log Q \le \frac{1}{2} (\log 2W + y_c)$
 and $2 y_c$ for $\frac{1}{2} (\log 2W+ y_c) < \log Q < \log 2W$.
It follows:
\ba
& & \bar n_c(Q\ge 2;\alpha) =
\frac{1}{(\log 2)^a (\log W)^A \bar N(W)}
\biggl\{ (y_c - \log 2W) \frac{A}{2(A+a-1)} \nonumber \\
& &\left[ \left(\frac{\log 2W + y_c }{2} \right)^{A+a-1} -
\left( \frac{\log 2W- y_c}{2} \right)^{A+a-1}  \right] +  \nonumber \\
&+& \frac{A}{A+a} \left[ \left( \frac{\log 2W + y_c}{2} \right)^{A+a} -
\left(\frac{\log 2W - y_c }{2} \right)^{A+a} \right] + \\
&+& y_c \frac{A}{A+a-1} \left[ \left( \log W \right)^{A+a-1} -
  \left( \frac{\log 2W + y_c }{2} \right)^{A+a-1} \right] \biggr\}
\nonumber
\ea
for $y_c < \log W/2$
and
\ba
& & \bar n_c(Q\ge 2;\alpha) =
\frac{1}{(\log 2)^a (\log W)^A \bar N(W)}
\biggl\{ (y_c - \log 2W) \frac{A}{2(A+a-1)} \nonumber \\
& & \left[  \left( \log W \right)^{A+a-1} -
\left( \log 2 \right)^{A+a-1}  \right] +  \nonumber \\
&+& \frac{A}{A+a} \left[ \left( \log W \right)^{A+a} -
\left( \log 2 \right)^{A+a}  \right] \biggr\}
\ea
for $ y_c \ge \log W/2$.

\begin{figure}[p]
\epsffile[100 100 500 500]{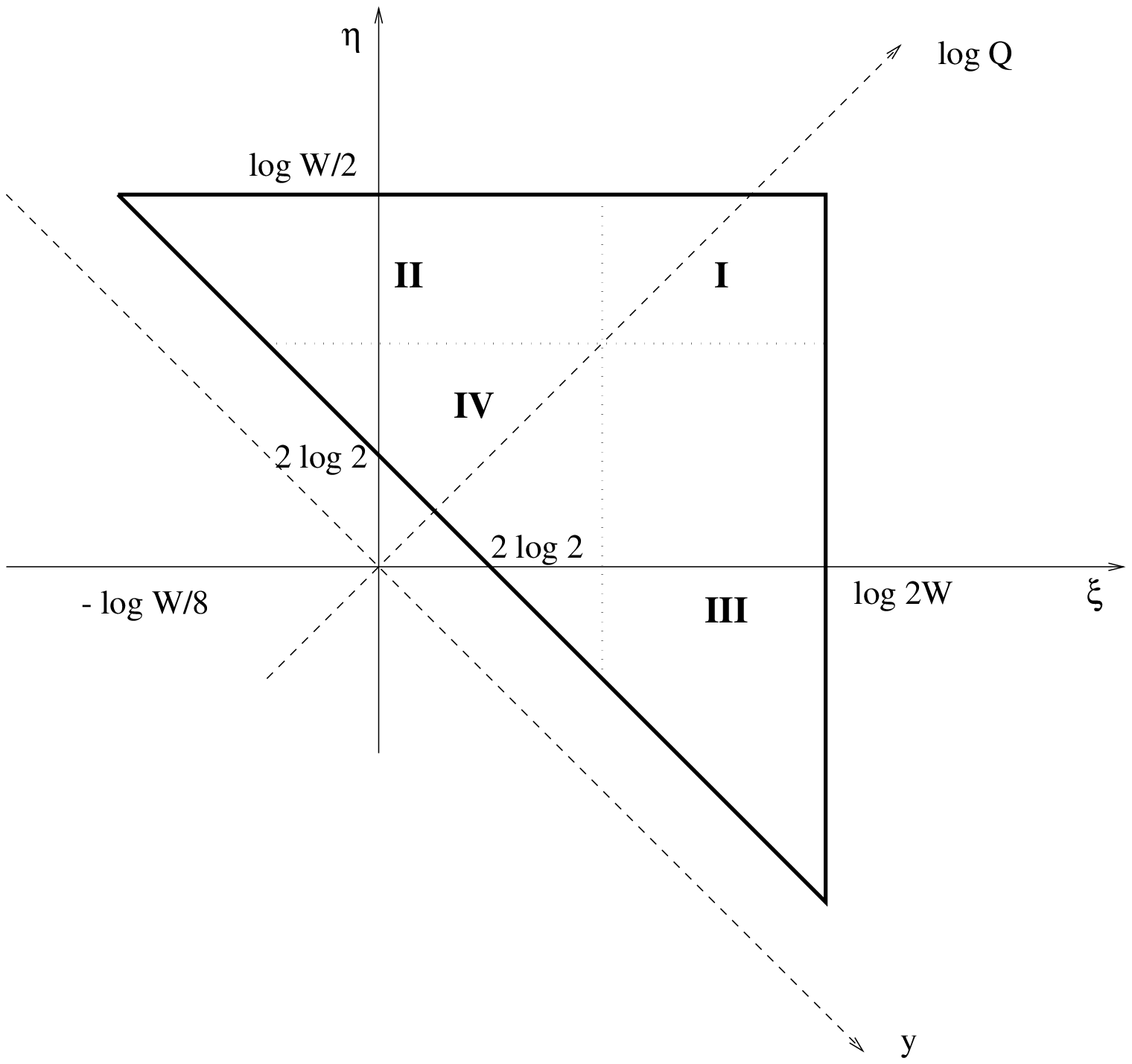}
\caption{Integration domain in the ($\xi,\eta$) plane
for $\bar n_c(Q \ge 2,\alpha)$ (see text, Eq.~\protect\eref{ex87}).
Dotted lines
indicate phase space domains where the integrand assumes different values.}
\label{triangolo}
\end{figure}

Now, we calculate the
second term, $\bar n_c(Q\ge 2;\beta)$. It is convenient to introduce
a new set of variables, i.e.:
\ba
\xi &=& \log Q + y  \\
\eta &=& \log Q - y  \; .
\ea
In these variables the initial integration domain turns out to be:
\be
\eta \le \log (\frac{W}{2}) \; ,\quad  \xi \le \log (2W) \; ,\quad
\eta + \xi > 2 \log 2 \; ,
\label{ex87}
\ee
as indicated in Figure~\eref{triangolo} (solid line).

\noindent
Notice also that the expression of the maximum contained in
Eq.~\eref{nfurry:more2} assumes different values in different sectors of
the  integration domain.

For $y_c$ in the interval
($\log 2 \div \log W/2$) one can identify four sectors (see
Figure~\eref{triangolo}, dotted lines):

\leftline{$I)$}
\be
y_c < \xi < \log 2  \qquad  y_c < \eta < \log W/2 \; ;
\ee
\leftline{$II)$}
\ba
& & - y_c + 2 \log 2 < \xi < y_c  \qquad
\qquad  y_c < \eta < \log W/2  \\
& &  \max(- y_c, - \log W/8) < \xi < -y_c+2 \log 2  \qquad
\qquad  -\xi+ 2 \log 2 < \eta < \log W/2 \; ; \nonumber
\ea
\leftline{$III)$}
\ba
& & y_c  < \xi < \log 2W  \qquad
\qquad  - y_c + 2 \log 2 < \eta < y_c  \\
& &  - \eta + 2 \log 2 < \xi < \log 2W  \qquad
\max(-y_c, -\log W/2) < \eta < - y_c + 2 \log 2 \; ; \nonumber
\ea
\leftline{$IV)$}
\be
 - y_c + 2 \log 2 < \xi < y_c  \qquad
\qquad  -\xi + 2 \log 2 < \eta < y_c \; ,
\ee
where $\pi_a(\Dy,Q,y)$ is:
\ba
&I)&  \qquad \qquad  2 y_c  \hfil \\
&II)&  \qquad \qquad  \xi + y_c  \hfil \\
&III)&  \qquad \qquad \eta + y_c   \hfil \\
&IV)&  \qquad \qquad  \xi + \eta  \hfil
\ea
respectively.

For $y_c$ values in the extreme allowed interval, $y_c \ge \log W/2$,
sectors $I)$ and $II)$ disappear and the other two sectors $III$ and $IV$
are modified as follows:

\medskip
\leftline{$III)$}
\be
\qquad y_c  < \xi < \log 2W  \qquad
\qquad  - \xi + 2 \log 2 < \eta < \log W/2 \; ;
\ee
\leftline{$IV)$}
\be
\qquad -  \log W/8 < \xi < y_c  \qquad
\qquad  -\xi + 2 \log 2 < \eta < \log W/2 \; .
\ee
Finally, for $y_c \le \log 2$, sector $IV)$ only disappears and the
others sectors become:

\medskip
\leftline{$I)$}
\ba
& &  \qquad y_c  < \xi < \log 2W  \qquad
- y_c + 2 \log 2 < \eta < \log W/2  \\
& &  - \eta + 2 \log 2 < \xi < \log 2W  \qquad
\quad  y_c < \eta < - y_c + 2 \log 2 \; ; \nonumber
\ea
\leftline{$II)$}
\be
\qquad -  y_c < \xi < y_c  \qquad
\qquad  -\xi + 2 \log 2 < \eta < \log W/2 \hfil \; ;
\ee
\leftline{$III)$}
\be
\qquad - \eta + 2 \log 2  < \xi < \log 2W  \qquad
\qquad  - y_c < \eta < y_c \hfil  \; .
\ee
The final analytical result
for $\bar n_c(Q\ge 2; \beta)$ can then be obtained.
In order to have an impression of the structure of this formula,
we show it explicitly
in the domain $\log 2 \le y_c \le \log W$:

\def\yc{{y_c}}  
\def\lW{(\log W)}
\def\la{(\log 2)}

\ba
& &\bar n_c(Q \ge 2,\beta)  = \frac{1}{1+ A \log (\frac{\lW}{\la})}
\Biggl\{
4\,\yc + {{7\,\yc}\over {2\,\lW}} +
  {{17\,\yc}\over {12\,\la}} + \nonumber \\
&+&  {{29\,\la\,\yc}\over {12\,{{\lW}^2}}} -
  {{2\,{{\la}^2}\,\yc}\over {3\,{{\lW}^2}}} +
  {{7\,{{\yc}^3}}\over {12\,{{\lW}^2}\,\la}} + \nonumber \\ \nonumber
&+&  {{4\,\lW\,\yc\,\log (\lW)}\over {3\,{{\la}^2}}} +
  {{2\,\yc\,\log (\lW)}\over {\la}} + \\ \nonumber
&+&
  {{2\,\bigl( -2\,{{\lW}^3}\,\yc +
        3\,{{\lW}^2}\,\la\,\yc - {{\la}^3}\,\yc \bigr)
\,\log (\lW - \la)}\over
    {3\,{{\lW}^2}\,{{\la}^2}}} + \\ \nonumber
&+&
  {{{{\bigl( -\lW + \yc \bigr) }^2}\,
\, \log \bigl(\lW - \yc \bigr)}
\over {6\, {{\lW}^2}\,{{\la}^2}}} \biggl[
  -{{\lW}^2} - 4\,\lW\,\la + \\ \nonumber
&-& 6\,{{\la}^2}  +
        2\,\lW\,\yc + 4\,\la\,\yc - {{\yc}^2} \biggr] + \\
&+&
  {{{{\left( \lW - \la - \yc \right) }^3}\,
      \log (\lW - \la - \yc)}\over
    {12\,{{\lW}^2}\,\la}} + \\ \nonumber
&+&
  {{{{\bigl( -\lW - \la + \yc \bigr) }^2}\,
      \, \log (\lW + \la - \yc)} \over
     {12\,{{\lW}^2}\,{{\la}^2}}}  \biggl[
       2\,{{\lW}^2} +  \\ \nonumber
&+& 15\,\lW\,\la  - 5\,{{\la}^2}
       -  4\,\lW\,\yc - 3\,\la\,\yc + 2\,{{\yc}^2} \biggr] + \\ \nonumber
&+&
  {{{{\left( \lW + \yc \right) }^2}\,
 \, \log (\lW + \yc)} \over {6 \, {{\lW}^2}\,{{\la}^2}}} \biggl[
       {{\lW}^2} + 4\,\lW\,\la + \\ \nonumber
&+& 6\,{{\la}^2}  +
        2\,\lW\,\yc + 4\,\la\,\yc + {{\yc}^2} \biggr] +
\\ \nonumber
&+&
  {{{{\left( -\lW + \la - \yc \right) }^3}\,
      \log (\lW - \la + \yc)}\over
    {12\,{{\lW}^2}\,\la}} + \\ \nonumber
&+&
  {{{{\left( \lW + \la + \yc \right) }^2}\, \log (\lW + \la + \yc)} \over
    {12\,{{\lW}^2}\,{{\la}^2}}}
      \biggl[ -2\,{{\lW}^2} + \\ \nonumber
&-& 15\,\lW\,\la + 5\,{{\la}^2} -
        4\,\lW\,\yc - 3\,\la\,\yc -
        2\,{{\yc}^2} \biggr]  \Biggr\} \; .
\ea

Having now integrated, as proposed in Eq.~\eref{nmedio:master}, the
product of the average parton multiplicity in a rapidity interval $\Dy$
for a single clan of given virtuality $Q$ and rapidity $y$, times the
probability that a parton of maximum allowed virtuality $W$ produces a
clan of virtuality $Q$ and rapidity $y$, by dividing the integration domain in
several regions, one obtains the wanted
final analytical expression for the average number of partons inside an
average clan in the interval $\Dy$ by assembling the various pieces as follows:
\ba
\bar n_c(\Dy,W)  &=& \bar n_c(Q<2) + \bar n_c(Q\ge 2) \nonumber \\
&=& \bar n_c(Q<2) + \bar n_c(Q\ge 2;\alpha)  + \bar n_c(Q\ge 2;\beta) \; .
\ea
The average number of partons in the shower, $\bar n(\Dy,W)$, can then be
calculated from Eq.~\eref{ex31}.

\section{Results: comments and discussion}

The motivation of this  work -- as pointed out in the Introduction -- has been
  to justify from first principle
clan structure
analysis in multiparticle production and to provide a sound theoretical basis
to the regularities discovered by its application to experimental data.
In order to do that, one would need
QCD calculations of a bi-dimensional shower evolution
in a region where  QCD predictions are not available and
calculations are very hard to be done.
One should therefore fully  rely on models. Our program
was to study analytically clan structure
in rapidity and virtuality variables
in a  parton shower model based on essentials of QCD in a
correct kinematical framework, the GSPS model.
Results of  analytical calculations     of the average number of clans
behavior  in virtuality and rapidity variables were reported in \cite{GSPS}
 and summarized in the introduction.  In order to complete
our initial program we propose in this paper
to discuss the properties  of the second
parameter of clan structure analysis, i.e., of the average number of
partons per clan, in rapidity and virtuality variables
as resulting from the analytical calculations
presented in  previous Sections.

Notice that these calculations have
been performed by following very  mild    mathematical
assumptions and by choosing  for the parameters $A$ and $a$ of the GSPS
model the  values 2 and 1   respectively. It should be remembered
that $A$ and $a$  control the "length" of the parton
shower  and of clans  in the shower. The suggested choices avoid  nasty
inessential  calculations and   make possible  the analytical
solution of the model without hurting its logic.
It should be pointed out that  MC  calculations  based on the same
architecture of the GSPS model allow to predict   the behavior of
$\bar n_c(\Delta y,W)$ as well as  that of the average number of clans in
the same variables, $\bar N(\Dy,W)$,  for all values of the parameters
$a$ and $A$.
In addition MC calculations here can be considered as exact , i.e., the
result is obtained without any of the above mentioned mild
mathematical approximations used in the analytical calculation in order
to get  the explicit dependence
of $\bar N(\Delta y, W)$ and  $\bar n_c(\Delta y, W)$ on $\Dy$ and $W$.
This fact should be considered important for estimating the goodness of the
assumptions performed in the analytical calculations.

Knowing from previous work\cite{GSPS} the rapidity and virtuality dependence
of the average number of clan, the analytical calculation of the average number
of partons per clan in the same variables can be done in terms of the average
number of partons in the full  shower. This is indeed a quantity which - as
previously discussed - does not depend on the redefinition procedure
and therefore can safely be used to our purpose.

\begin{figure}[p]
\epsffile[100 100 500 500]{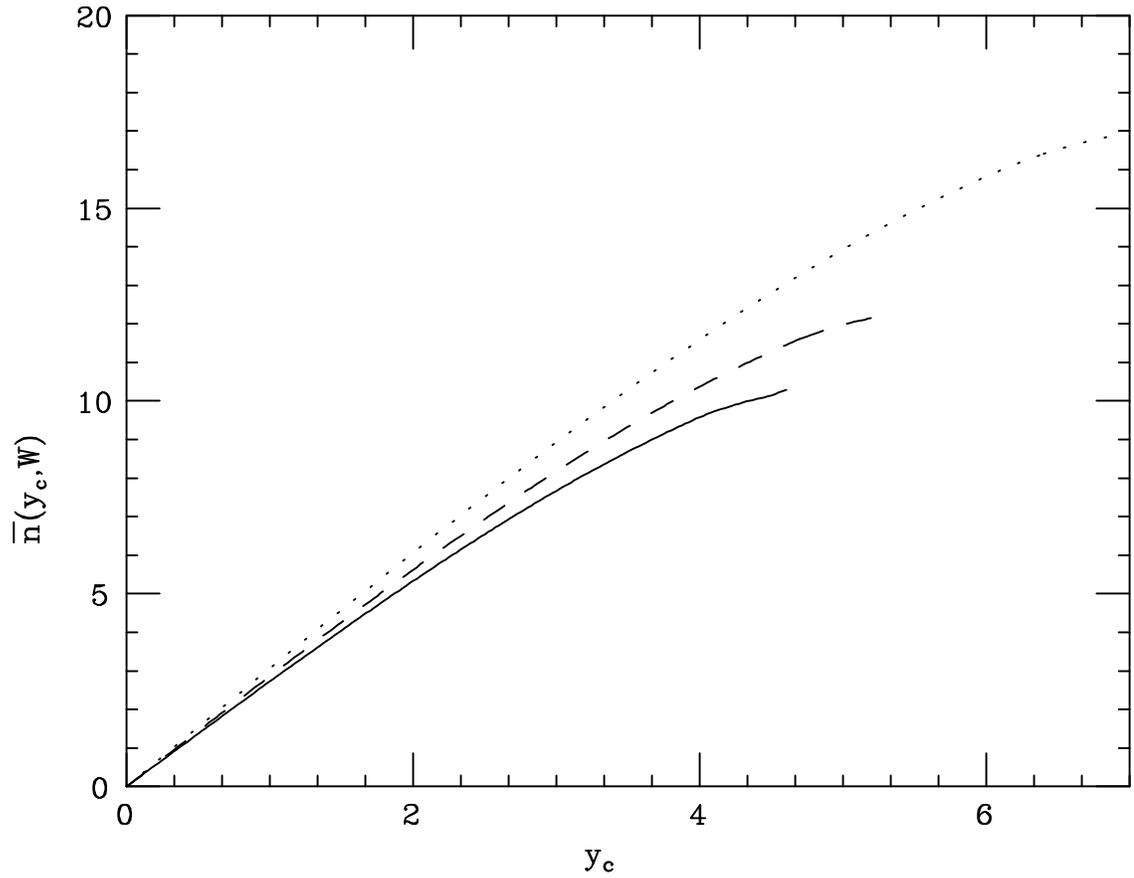}
\caption{Average number of partons in the shower, $\bar n(y_c,W)$,
 as a function of the width of the
rapidity interval $y_c$ obtained analytically
in the GSPS model with $A$ = 2, $a$ = 1 at different maximum allowed
virtualities $W$ = 50 GeV (solid line), 100 GeV (dashed line) and 500 GeV
(dotted line) (see Eq.~\protect\eref{ex26}).}
\label{avgn}
\end{figure}

The result of the analytical calculation   of the average number of
partons in the shower, $\bar n(\Dy,W)$,
as a function of the rapidity interval $\Dy$ and
of the maximum allowed virtuality $W$ with $A$=2 and $a$=1, Eq.~\eref{ex26},
is shown in Figure~\eref{avgn} for $W$= 50 GeV (solid line), $W$=100 GeV
(dashed line) and $W$= 500 GeV (dotted line).
The average  parton multiplicity grows   almost linearly
with rapidity for relatively small $\Dy$ intervals and then it is slowly
bending for $\Dy$ intervals approaching to fps, where it reaches its
maximum ($\bar n({\mathrm{fps}}) \simeq 10$ for  $W$ = 50 GeV,
$\bar n({\mathrm{fps}}) \simeq 12$
for $W$ =  100 GeV  and $\bar n({\mathrm{fps}})
\simeq 17$  for $W$ = 500 GeV).

\begin{figure}[p]
\epsffile[100 100 500 500]{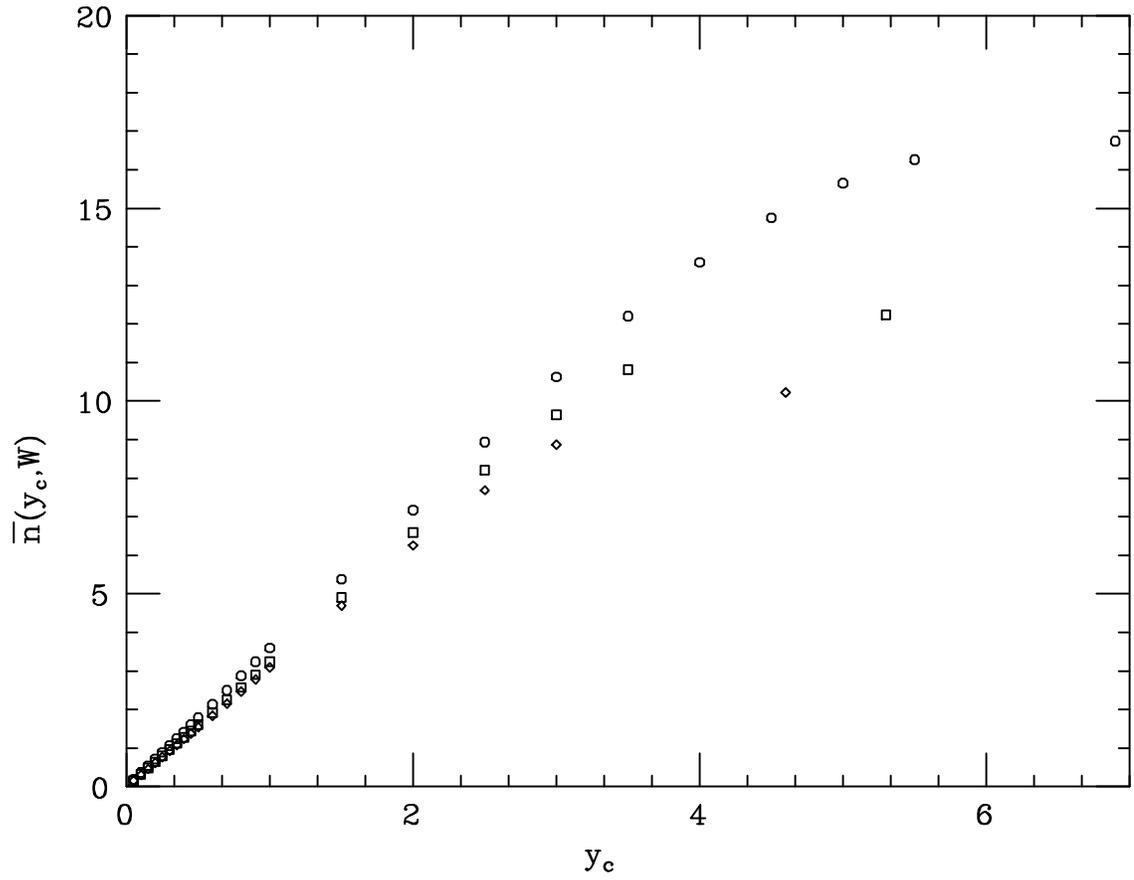}
\caption{Same as in Figure~\protect\eref{avgn},
but for the Monte Carlo version of the model at different maximum allowed
virtualities $W$ = 50 GeV (diamonds),
100 GeV (squares) and 500 GeV (circles).}
\label{avgn:MC}
\end{figure}

Monte Carlo calculations of $\bar n(y_c,W)$ for the same $W$ values and for the
same choice of the parameters $A$ and $a$ are given in Figure~\eref{avgn:MC}.
The general trend is the same as that
in Figure~\eref{avgn}.
The agreement between the two calculations is very good in fps.
For smaller  $\Dy$  intervals Monte Carlo calculations are sistematically
higher into few percent and allow to estimate the amount of the  approximations
done in the analytical calculation.

\begin{figure}[p]
\epsffile[100 100 500 500]{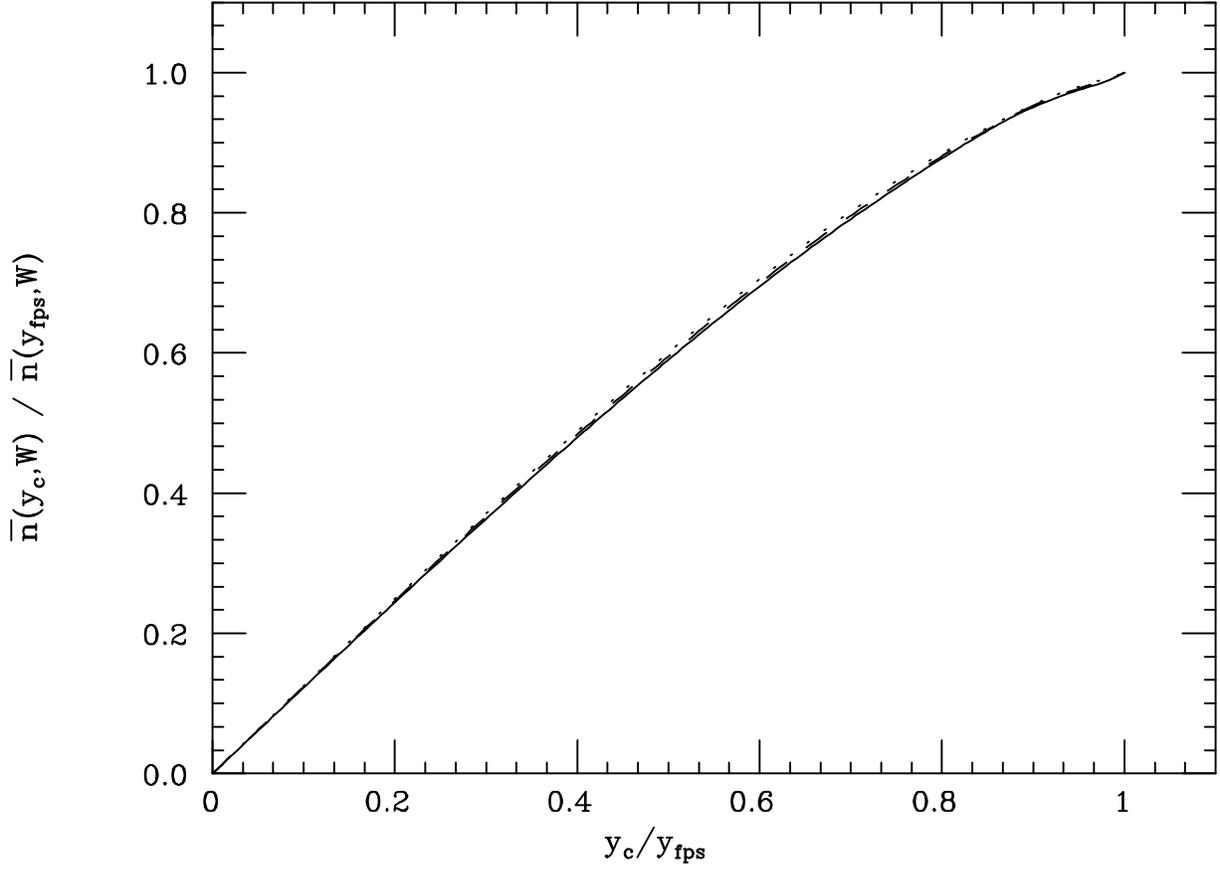}
\caption{Normalized average number of partons in the shower,
$\bar n(y_c,W)/\bar n({\mathrm{fps}},W)$ as a function of rescaled
rapidity interval, $y_c/{\mathrm{fps}}$ for $A$ = 2, $a$ = 1 at
different maximum allowed virtualities
$W$ = 50 GeV (solid line), 100 GeV (dashed line) and 500 GeV (dotted line).}
\label{rescaled}
\end{figure}

It is interesting to remark that the normalized average number of partons in
the shower, $\bar n(\Delta y, W)/\bar n({\mathrm{fps}}, W)$
scales in virtuality  as a function
of the rescaled rapidity interval, $\Dy/{\mathrm{fps}}$,
see Figure~\eref{rescaled}.
This scaling in $W$ is found to depend on the parameter $a$, as different
values of $a$ give different scaling curves,
differently from the scaling found in \cite{GSPS}
for $\bar N(\Dy,W)/\bar N({\mathrm{fps}},W)$ which is independent
of the mechanism inside clans.

\begin{figure}[p]
\epsffile[100 100 500 500]{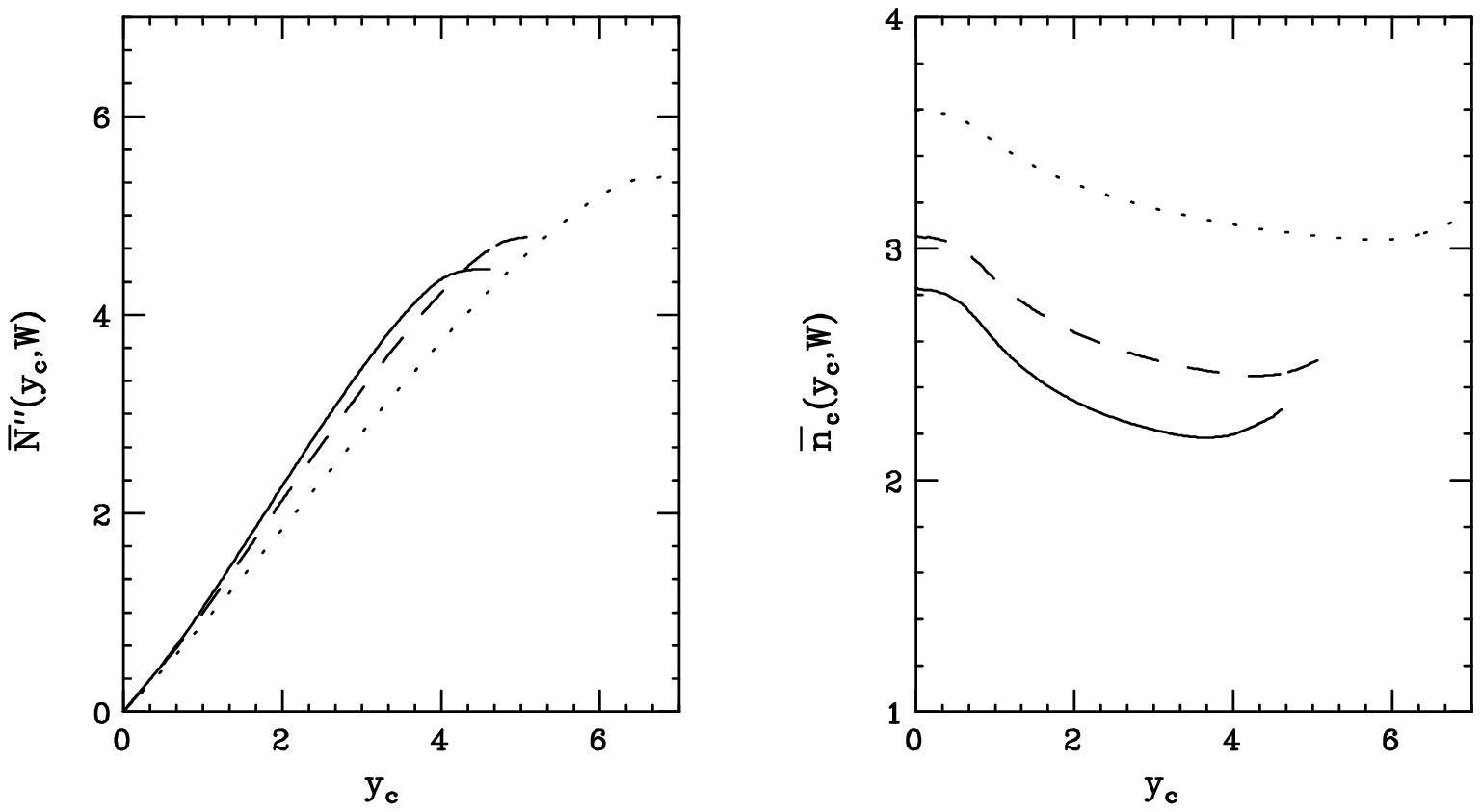}
\caption{{\bf a)}:
average number of clans, $\bar N''(y_c,W)$,
defined by binomial convolution as in Eq.~\protect\eref{nmediosecondo}, as a
function of the width of the
rapidity interval $y_c$ at different maximum allowed virtualities
$W$ = 50 GeV (solid line), 100 GeV
(dashed line)  and 500 GeV (dotted line).
Analytical solution with $A$ = 2. The solution is of course independent
of $a$.
{\bf b)}: corresponding
average number of partons per clan, $\bar n_c(y_c,W)$, as a
function  of the width of the
rapidity interval $y_c$ at
different maximum allowed virtualities
$W$ = 50 GeV (solid line), 100
GeV (dashed line) and 500 GeV (dotted line).
Analytical solution with  $A$ = 2 and $a$ = 1.
}
\label{noanc}
\end{figure}

Figure~(\ref{noanc}a) shows
the behavior of the average number of clans, $\bar N'(\Dy,W)$, in
showers of maximum allowed virtuality $W$ equal to 50 GeV (solid line),
100 GeV (dashed line) and 500 GeV (dotted line)  as a function
of rapidity interval $\Dy$ (Eq.~\eref{nmediosecondo}).
The analytical calculation has
been done  with $A$=2 . Being the first step    only of the
production process involved in the calculation, the result does not
depend  of course   on the parameter $a$.
The Figure content
is the same  discussed in  Figure~8 of \cite{GSPS}.

Having shown in Figures~\eref{avgn} and ~(\ref{noanc}a)
the general trends of the average number of
partons in the shower and of the average number of clans  as  functions
of rapidity and virtuality variables  predicted by  the analytical
calculations according to Section~4 and our previous work\cite{GSPS},
we are now ready to calculate
the average number of partons per clan in the same variables, i.e.,
$\bar n_c(\Delta y, W)$, simply  by forming the ratio of the
above two quantities.

\begin{figure}[p]
\epsffile[100 100 500 500]{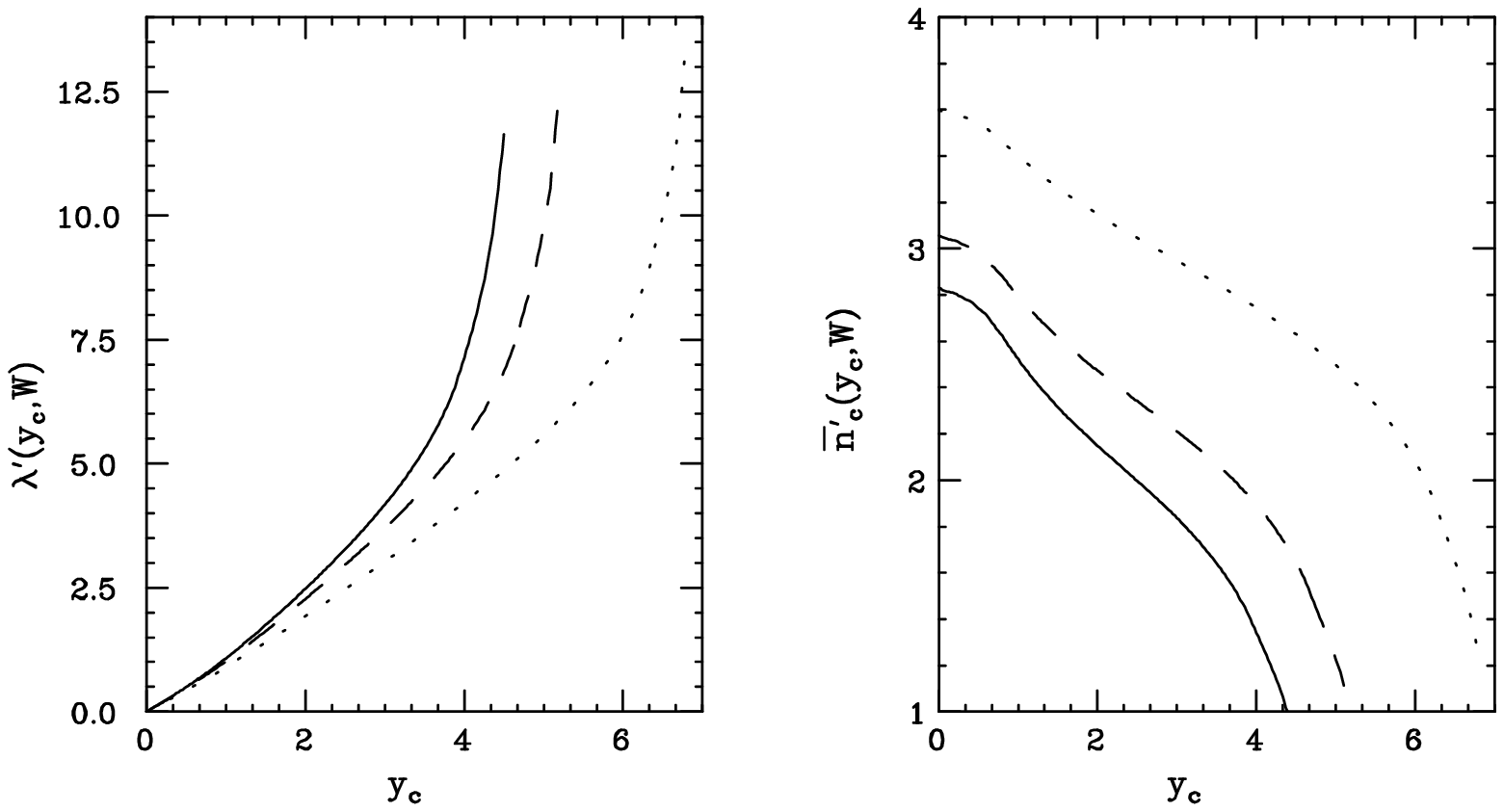}
\caption{{\bf a)}: average number of clans, $\lambda'(y_c,W)$,
defined by requiring that the total MD be
a CPD as in  Eq.~\protect\eref{lambdaprimo}, as a function
of the width of the
rapidity interval $y_c$ at
different maximum allowed virtualities $W$ = 50 GeV (solid line),
100 GeV (dashed line) and 500 GeV (dotted line).
Analytical solution with $A$ = 2. The solution is of course
independent of  $a$.
{\bf b)}: corresponding average number of partons per clan,
$\bar n_c(y_c,W)$, as a function of
the width of the rapidity interval $y_c$
at different maximum allowed virtualities
$W$ = 50 GeV (solid line), 100 GeV (dashed line) and 500 GeV
(dotted line). Analytical solution with $A$ = 2 and $a$ = 1.
}
\label{anc}
\end{figure}

$\bar n_c(\Delta y, W)$ values in rapidity intervals for $W$ = 50 GeV
(solid line),
$W$ = 100 GeV (dashed line) and $W$ = 500 GeV (dotted line),
with parameters $A$ and $a$
equal to 2 and 1 respectively,
are plotted in Figure~(\ref{noanc}b). The result is reasonable
although not fully satisfactory.
An anomalous behavior   is clearly visible
 for large rapidity intervals ($\bar n_c(\Delta y, W)$  increases
as  $\Dy \to {\mathrm{fps}}$ in contrast with what one would expect
from standard  clan structure analysis at parton level)
and
for small rapidity intervals  ($\bar n_c (\Delta y, W)$  is shown to point
to constant values, which differ for different $W$ values in the limit
$\Dy \to 0$  in contrast with the expected  $W$ independent $\bar n_c(0,W)=1$
value).
In order to cure the first anomaly we decided to use the definition
from the CPD form, Eq.~\eref{lambdaprimo}, which is relevant in fps,
and accordingly to use Eq.~\eref{ncprimo:anc} for the average number of
partons per clan.
The relevance of this change
is shown in Figure~(\ref{anc}a), where $\lambda'(\Dy,W)$ is shown
to diverge as $\Dy \to $ fps as expected; because of this behavior, however,
the  procedure cannot be extended to fps. It will be noticed
in Figure~(\ref{anc}b) that indeed the above mentioned anomaly  is removed:
the result becomes consistent with our scheme
by assuming that only clans with at least
one parton are allowed.

\begin{figure}[p]
\epsffile[100 100 500 500]{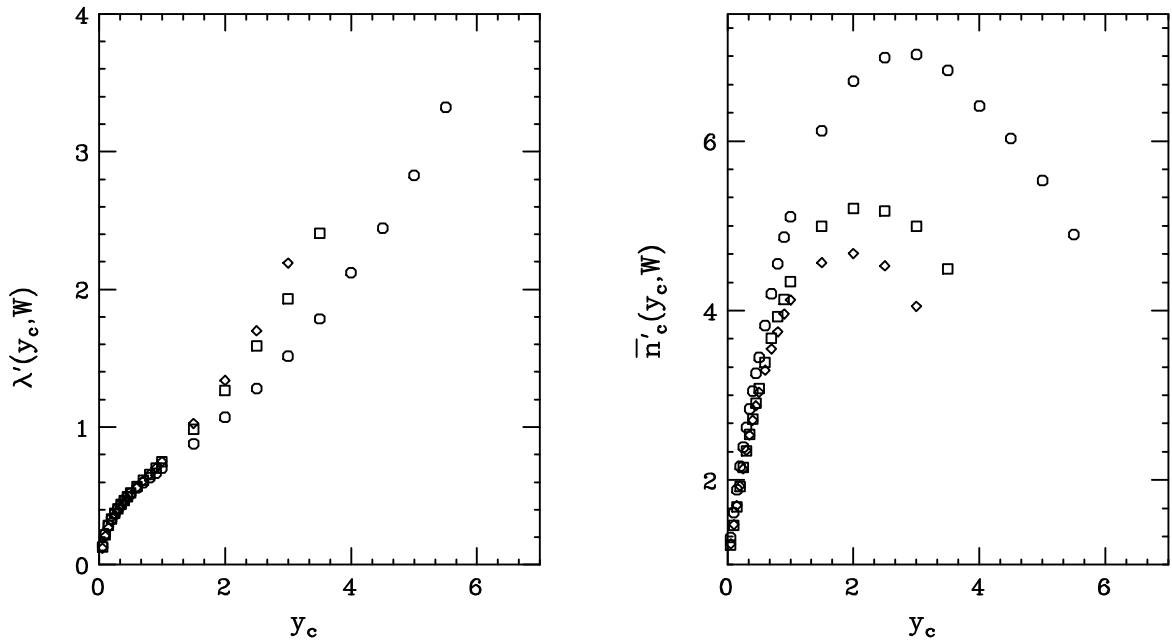}
\caption{Same as in Figure~\protect\eref{anc},
but Monte Carlo results at different maximum allowed virtualities
 $W$ = 50 GeV (diamonds), 100 GeV (squares) and 500 GeV (circles).}
\label{MC}
\end{figure}

The most natural   motivation of the appearence of the second anomaly
lies in our opinion in the approximations used   in  the analytical
calculations as already shown by comparing
Figures~\eref{avgn} and ~\eref{avgn:MC}.
The approximations  imply deviations from the exact behavior for
$\Dy <$ fps.
In order to check this guess we propose
to calculate the average number of clans and the average number of partons
per clan with the MC.
Results of the MC calculations are shown in
Figures~(\ref{MC}a) and ~(\ref{MC}b).
Their trends fully coincide with standard results already
observed in clan structure analysis applied to parton showers in quark and
gluon jets.
The predictive power of the model under investigation
is therefore confirmed. In addition,
for small rapidity intervals the expected  $\bar n_c(\Delta y = 0,W)=1$
limit is recovered.

We have shown that the  parton shower model, which we
built by assuming  QCD-inspired dependence of the splitting
functions in virtuality and in rapidity
 and Sudakov form factors for their normalization
("essentials  of QCD"), and by allowing at each step in the cascading
local violations of
energy-momentum conservation law, but requiring its global validity,
has an extraordinary predictive power  in  regions not accessible to
full perturbative QCD. In particular, we performed analytical calculations
of   the rapidity and virtuality dependence of the average number
of clans, of the average parton multiplicity  and
of the average number of partons inside a clan in a parton shower
initiated by an ancestor parton of given virtuality.
Clans here should be intended of course in a more general sense than that
usually referred to in the literature and  linked hystorically to NB
regularity: they are the natural language for describing a two step
process in terms of a CPD and can be identified in our parton shower model
with effective intermediate independent parton (gluon) sources
in the  shower.
Results of our analytical calculations  are consistent with what we know
on clan properties  in single quark and gluon jets disentangled
by using  a jet finding algorithm  and analyzed   at parton level
by assuming  the  validity of GLPHD.
In addition it should be pointed out that, even if
the change of $A$ and $a$ parameters
do not influence  the qualitative behavior of
the average number of clans and
the average number of partons per clan  in rapidity and virtuality,
it actually affects their numerical values.
 Since $A$ and $a$ parameters control
the length of the parton shower and the development of parton
cascading inside a clan,
would one describe  a  two parton species  shower initiated by
an ancestor gluon or quark, this
flexibility of the model could be very useful in order to
describe jets of different origin. This mechanism could be
also very helpful in a more general framework for explaining the
observed different behavior of clans   in different classes of
high energy collisions.
In summary,  these findings suggest that the dynamics of multiparticle
production is controlled by a QCD inspired two step process
dominated by independent intermediate parton (gluon) sources, which
we call clans,
and  by their cascading. The features which make the model appealing
are its simple structure, the
possibility to calculate  in its framework
analytically important physical quantities
and its flexibility.

\section{Acknowledgement}

One of us (S.L.) would like to thank D.A.A.D. for financial support.

\vfill\eject

\end{document}